\documentstyle[epsf,pre,aps]{revtex}
\begin{document}
\twocolumn[\hsize\textwidth\columnwidth\hsize\csname @twocolumnfalse\endcsname
\draft
\title{Monte Carlo Simulation of Smectic Liquid Crystals and the Electroclinic
Effect: the Role of the Molecular Shape}
\author{Jianling Xu and Robin L. B. Selinger\thanks{Author to whom
correspondence should be addressed.}}
\address{Physics Department,
Catholic University of America,
Washington, DC 20064}
\author{Jonathan V. Selinger, B. R. Ratna, and R. Shashidhar}
\address{Center for Bio/Molecular Science and Engineering,
Naval Research Laboratory, Code 6900, \\
4555 Overlook Avenue, SW,
Washington, DC  20375}
\date{July 27, 1999}
\maketitle

\begin{abstract}
Using Monte Carlo simulation methods, we explore the role of molecular shape in
the phase behavior of liquid crystals and the electroclinic effect.  We study a
``bent-rod'' mesogen shaped like the letter Z, composed of seven soft spheres
bonded rigidly together with no intra-molecular degrees of freedom.  For
strongly angled molecules, we find that steric repulsion alone provides the
driving force for a smectic-C phase, even without intermolecular dipole-dipole
interactions.  For weakly angled (nearly rod-like) molecules, we find a stable
smectic-A (SmA) phase and a strong electroclinic effect with a saturation tilt
angle of about $19^\circ$.  In the SmA phase we find evidence of vortex-like
point defects.  We also observe a field-induced nematic-smectic phase
transition.
\end{abstract}


\vskip2pc]
\narrowtext

\section{Introduction}

The response of smectic liquid crystals to applied electric fields has been
extensively studied for both basic research and applications.  One subject of
particular interest is the electroclinic effect, which occurs in the smectic-A
(SmA) phase of chiral molecules.  In the electroclinic effect, an applied
electric field in the smectic layer plane induces a tilt of the molecules
relative to the layer normal, in a direction orthogonal to the field.  The
magnitude of the induced tilt scales linearly with the applied electric field
for low fields, and then saturates at higher fields.  This effect was predicted
by Meyer on the basis of symmetry~\cite{meyer}, and it was subsequently
observed experimentally by Garoff and Meyer~\cite{garoff}.  It is now being
exploited for electro-optic devices that display a continuous gray scale as a
function of applied electric field, such as spatial light
modulators~\cite{collings,andersson}.

To optimize electroclinic liquid crystals for device development, one needs a
theoretical understanding of how the electroclinic tilt depends on electric
field, temperature, and molecular structure.  So far, most theoretical work on
the electroclinic effect has been through Landau theory, i.e.\ a minimization
of the free energy expanded in powers of the molecular tilt and the
electrostatic polarization~\cite{meyer,blinc,moddel}.  This work explains
certain aspects of the electroclinic effect---in particular, it shows how the
tilt and polarization depend on field for low fields, and it shows how the
susceptibility to a field increases as the system approaches the second-order
phase transition from the SmA to the smectic-C (SmC) phase.  However, some
important questions about the electroclinic effect are not addressed by Landau
theory.  The first and most general question is:  How sensitive is the
electroclinic effect to molecular shape?  In other words, how much does the
electroclinic susceptibility change with slight details of molecular structure?
A second and more specific question is:  How does the applied electric field
change the distribution of molecular orientations?  Does it make the molecules
tilt as rigid rods from an initially untilted state to a tilted state?  Or does
it change a state of disordered tilt in random directions into a state of
ordered tilt in one direction?  The latter alternative is suggested by the de
Vries description of the SmA phase~\cite{devries}.

To address these questions, in this paper we present a series of Monte Carlo
simulations of smectic liquid crystals.  Simulation is an appropriate tool to
address these questions for two reasons.  First, in simulations we can begin
with a microscopic model for the molecular structure and determine the
large-scale order of the liquid-crystal system as a function of thermodynamic
variables such as temperature, density, and applied field.  We can then make
small changes in the molecular shape and see how these changes affect the
large-scale order of the system.  Thus, we can determine how macroscopic
properties such as the electroclinic susceptibility depend on details of the
molecular shape.  Second, in simulations we can take snapshots of the
positions and orientations of all the molecules in the system, and hence can
extract any correlation function to characterize the system.  This information
is not available in Landau theory, and is generally difficult to extract from
experiments.  Hence, simulations give us new information about the distribution
of molecular orientations as a function of electric field, and about
topological defects in the molecular orientations.

In these simulations, we use a ``bent-rod'' rigid molecule with the oblique
shape shown in Fig.~1.  This shape is motivated by three considerations.
First, the three-dimensional structure of many liquid-crystal molecules, such
as the homologous series KNnm, has this general shape~\cite{kn125}.  In the
center is a rigid molecular core, which defines the optical axis of the
molecule, and on both ends are hydrocarbon chains, which extend out at an angle
from the core.  In the homologous series KNnm, the electroclinic tilt angle of
the SmA phase can be increased by making the hydrocarbon chains longer, thus
making the molecules more oblique.  Second, the Boulder model for ferroelectric
liquid crystals shows that molecules in the SmC phase typically take the shape
of bent cylinders~\cite{boulder}.  For that reason, we can regard this shape as
a generic feature of smectic liquid crystals.  Third, density functional theory
has been used to predict the phase diagram of parallel offset hard cylinders, a
shape similar to bent rods~\cite{tarazona}.  That work showed a high-density
SmC phase for molecules with a higher offset ratio, i.e. the more oblique.
These results confirm that the obliqueness of molecular shape is an important
parameter to determine the phase behavior of smectic liquid crystals.

To simulate a simple molecular structure with a bent-rod shape of variable
obliqueness, we use a molecule composed of seven spheres arranged in the shape
of the letter Z, as illustrated in Fig.~1.  The spheres are ``glued'' rigidly
together with no intra-molecular degrees of freedom, with a bend angle $\theta$
between the core and tail portions of the molecule.  We consider the cases
$\theta=45^\circ$, which is quite oblique, and $\theta=5^\circ$, which
approaches the rod-shaped limit of $\theta=0^\circ$. Each molecule also has a
dipole moment that lies perpendicular to the molecular backbone, as shown,
giving the molecule a chiral structure.  The molecules interact through a soft
repulsive sphere-sphere pair potential, and each molecular dipole interacts
with the applied electric field.  We neglect dipole-dipole interactions as an
approximation to simplify the computations.

These simulations provide clear evidence that steric repulsion alone can give
rise to order in the molecular tilt, even without including intermolecular
dipole-dipole interactions.  Furthermore, they show that the bend angle
$\theta$ plays a major role in determining phase behavior.  For
$\theta=45^\circ$ the system has a phase transition directly from the isotropic
phase to the SmC phase.  By contrast, for $\theta=5^\circ$, the system has
nematic and SmA phases, each stable over a wide range of temperature.  In the
absence of an applied electric field, the molecules of the SmA phase are not
aligned with the layer normal but rather are tilted in random directions, and
the orientation of the tilt exhibits vortex-like point defects.  When an
electric field is applied, the magnitude of the molecular tilt increases and
the direction of the tilt becomes more ordered, giving a strong electroclinic
effect.  At high fields, the electroclinic tilt angle saturates at
approximately $19^\circ$.  The simulations also show that a high electric field
applied to the nematic phase induces a transition into the SmA phase, showing
another ordering effect of the field.

The plan of this paper is as follows.  In Sec. II we describe the details of
the model and the computational method that we used.  In Sec. III we present
the results of the simulations for bend angle $\theta=45^\circ$ and
$\theta=5^\circ$.  In particular, we show the electroclinic effect in the SmA
phase for $\theta=5^\circ$.  Finally, in Sec. IV, we discuss the significance
of these results for experiments on smectic liquid crystals.

\section{Simulation Method}

In our simulations, we consider molecules composed of seven soft spheres
arranged in the rigid bent structure shown in Fig.~1.  The molecular director
is defined as the unit vector along the five-sphere core of the molecule.  The
interaction between molecules is reduced to an interaction between different
spheres in different molecules.  Intramolecular interactions and degrees of
freedom are suppressed.  The sphere-sphere interaction potential is the
truncated Lennard-Jones potential, also known as the Weeks-Chandler-Anderson
potential~\cite{wca}, cut off at its minimum so there is no attractive tail:
\begin{equation}
U^{\rm int}_{mn}=
\cases{\displaystyle 4\epsilon\left[
\left({\sigma\over r_{mn}}\right)^{12}-
\left({\sigma\over r_{mn}}\right)^{6}
\right]+\epsilon,
&if $r_{mn}\le r_{c}=2^{1/6}\sigma$;\cr
0,&otherwise.\cr}
\end{equation}
where $r_{mn}=|\vec{r}_{m}-\vec{r}_{n}|$ and $m$ and $n$ are the sphere indices
in different molecules.  We choose this short-range repulsive interaction to
reduce required computation time and to focus on the role of steric effects
without any contribution from attractive interactions.  For the rest of this
paper, we measure lengths in units of $\sigma$ and energies in units of
$\epsilon$.  In addition, each molecule interacts with the applied electric
field $\vec{E}$ through the coupling
\begin{equation}
U^{\rm dipole}_j = -{\vec E}\cdot\vec{p}_j,
\end{equation}
where $\vec{p}_j$ is the dipole moment of molecule $j$.  The molecular dipole
moment is defined to have unit magnitude, which gives a scale for the electric
field.  We simulate 500 molecules in a flexible three-dimensional box with
periodic boundary conditions.  We keep the system with constant volume density
0.75 Lennard-Jones particles per unit volume, and allow the aspect ratio of the
simulation cell to adjust according to the Metropolis algorithm.  We do not
allow the cell to shear.

The system is prepared by a procedure analogous to the experimental technique
of cooling in a strong aligning field to avoid the formation of smectic
domains.  We begin the simulations at the high temperature $k_B T=20.0$, with
the box size of $11.5\times 11.5\times 35.0$.  This aspect ratio favors the
formation of a five-layered smectic phase.  In the initial state, the molecules
have random positions but all the molecules are ``double-aligned,'' that is,
both the directors and the dipole moments are aligned.  During the preliminary
cooling procedure, we suppress all orientational degrees of freedom and allow
the molecules to diffuse while remaining double-aligned.  The temperature of
the system is reduced slowly at a rate $10^{-4}$ per Monte Carlo step.  The
system comes to equilibrium quickly.  In about 10,000 Monte Carlo steps, the
molecules form five distinct layers.  If the layer normal is not parallel to
the $z$ axis, we measure the angle away from the $z$ axis, adjust the director
of the molecules, choose a new random initial configuration, and repeat the
simulation to get a layered system with the layer normal along the $z$ axis,
with no defects in the layer structure.

Once we reach this double-aligned smectic state, we reduce the temperature to
about 1.5, still in the double-aligned state.  Then, after the system is
in equilibrium, we switch on the three rotational degrees of freedom for each
molecule and equilibrate for an additional 100,000 Monte Carlo steps per
particle.  In one Monte Carlo step, each randomly selected molecule attempts
three translations and three rotations.

To characterize the phase behavior of the system, we particularly use three
order parameters.  First, the nematic order tensor $Q$ represents the strength
and direction of orientational order of the molecules.  It is defined as
\begin{equation}
Q_{\alpha\beta}=\left\langle\frac{1}{N}\sum_{j=1}^{N}
\left(\frac{3}{2}n_{j\alpha}n_{j\beta}-\frac{1}{2}\delta_{\alpha\beta}\right)
\right\rangle,
\label{nematic}
\end{equation}
where $\vec{n}_j$ is the director along the core of molecule $j$ and $N=500$
is the number of molecules.  The eigenvector corresponding to the maximum
eigenvalue of $Q$ is the average director of the system.  If the eigenvalue is
1, the molecular directors are completely aligned; if the value is lower it
reflects less perfect alignment.  Second, the polarization $\vec{P}$ represents
the degree of orientation of the molecular dipole moments.  It is defined as
the vector average
\begin{equation}
\vec{P}=\left\langle\frac{1}{N}\sum_{j=1}^{N}\vec{p}_j\right\rangle.
\label{polardef}
\end{equation}
Third, the smectic order parameter $\sigma$ represents the strength of the
density modulation along the $z$ direction.  It is defined as
\begin{equation}
\sigma=\left\langle\frac{1}{N}\sum_{j=1}^N e^{2\pi i z_j /d}\right\rangle,
\label{smectic}
\end{equation}
where $z_j$ is the $z$-coordinate of the center of mass of molecule $j$ and $d$
is the smectic layer wavelength, which is one-fifth of the $z$-dimension of the
simulation cell.

\section{Results}

\subsection{Forty-Five Degree Bent-Rod Molecules}

For the molecules with the large bend angle $\theta=45^\circ$, the phase
sequence is crystal-SmC-isotropic.  The SmC phase is stable over a wide range
of temperature, from approximately $k_B T=0.5$ to 1.5 (in Lennard-Jones units).
A sample configuration is shown in Fig.~2.  The polarization of any single
layer, defined by Eq.~(\ref{polardef}), is $P=0.85$, indicating nearly
perfect orientational order.

In spite of the high orientational order within each layer, the local tilt
direction of each layer is only loosely coupled to that of adjacent layers, and
it tends to wander.  This type of behavior was evident also in the simulation
study of Affouard et al.~\cite{affouard} on a related system.  It is similar to
the proposed random smectic-C$_{\rm R}$ phase, which has been suggested as a
model for the thresholdless switching observed experimentally in certain
smectic liquid crystals~\cite{fukuda}.  An alternative model has recently been
proposed for these experiments~\cite{rudquist}, but the smectic-C$_{\rm R}$
phase remains a theoretical possibility for future materials.  Indeed, this
proposed phase with random orientations of adjacent layers can be viewed as one
version of the sliding phase that has been investigated in recent theoretical
work~\cite{ohern}.

One possible explanation for the low interlayer correlations in our simulations
is that there is very little interaction between the tilt directions in
adjacent layers, because the intermolecular potential is purely repulsive and
because there is hardly any interdigitation between the layers.  As a result,
the adjacent layers should have very little preference for synclinic
(ferroelectric) or anticlinic (antiferroelectric) order, and they should be
fairly free to wander between these extremes.  An alternative explanation is
that the layers might prefer anticlinic order, but they are frustrated because
the system has an odd number of layers (five).  It is interesting to note that
Affouard's simulation also included an odd number of layers (three).  This
latter explanation seems less likely, however, because the interaction between
layers does not seem to favor anticlinic order.

Note that we observe the SmC phase even though we have not included
dipole-dipole interactions in our intermolecular potential, indicating that
steric repulsion defined by molecular shape is sufficient to produce order in
the molecular tilt direction.  Electrostatic interactions are not required to
produce a tilted smectic~\cite{glaser}.  Presumably the inclusion of
dipole-dipole interactions in our simulation would increase the temperature
range over which the SmC is stable, and it would likely increase the coupling
between the tilt directions in adjacent layers.

\subsection{Five Degree Bent-Rod Molecules}

The molecule with bend angle $\theta=5^\circ$ looks very nearly like a rod, but
it has properties quite different from a purely rod-shaped molecule.  The
$\theta=5^\circ$ system has a stable SmA phase over a temperature range of
$k_B T = 0.7$ to 3.0.  A sample configuration of the SmA phase is shown in
Fig.~3a.  The smectic order parameter defined by Eq.~(\ref{smectic}) is very
high, about 0.9.

The molecules in each layer of the SmA phase have approximately zero average
tilt and no net polarization.  However, a close look at the structure of an
individual layer shows that the {\it local} molecular tilt is nonzero but that
defects cause the net tilt to vanish, as shown in Fig~3b.  In some
configurations, these point defects in the local tilt appear to be vortices
analogous to those seen in, for example, an $xy$ model~\cite{xy}.  Comparison
of defect structures in adjacent layers shows that there is no strong
correlation in defect location between layers, indicating that these defects
are truly point vortices and do not thread through all five layers of the
system.  In this respect, they are analogous to the ``pancake'' vortices seen
in layered superconductors with weak interlayer coupling~\cite{crabtree}.

When we apply an electric field in the SmA layer plane, the molecules tilt
showing a clear electroclinic effect.  The observed polarization responds
rapidly to the applied field, coming close to its equilibrium value in only
several thousand Monte Carlo steps, while the tilt angle takes up to 500,000
Monte Carlo steps to equilibrate.  This equilibration would likely have been
faster if we had implemented degrees of freedom that allowed shear deformation
of the simulation cell, but clearly it is much longer than the equilibration
time for the polarization.  Figure~4a shows the SmA phase under a strong
applied field $E=10$.  When we examine a layer from this system, we observe
that the vortex-like defects have vanished, and the molecules in the same layer
are all closely aligned, as shown in Fig.~4b.

We can compare the measured polarization response to the applied field with the
prediction of a simple spin model.  The molecules in a smectic layer are
localized with directors pointing in almost the same direction.  The most
active movement is the rotation of the molecular dipole moment around the
director.  In view of this property, we consider the molecules as
two-dimensional independent dipoles with only one effective rotational degree
of freedom.  The net polarization can then be written as
\begin{equation}
P=\frac{\int_0^{2\pi} p\cos\theta e^{E p\cos\theta/k_B T} d\theta}
{\int_0^{2\pi} e^{E p\cos\theta/k_B T} d\theta}
=\frac{I_1(E p/k_B T)}{I_0(E p/k_B T)}
\label{polar}
\end{equation}
where $I_0$ and $I_1$ are modified Bessel functions and $p=1$ is the magnitude
of the dipole moment of a single molecule.  The simulation results for
polarization vs.\ field are plotted together with the analytic prediction in
Fig.~5a, and are in close agreement.  Indeed, the agreement is much closer than
one would expect from such a simple model---one would expect the interactions
among the molecules to give collective order that would give a higher initial
slope to the the polarization vs.\ field curve.  This agreement shows that the
alignment of molecular dipole moments with the electric field is a
single-molecule effect rather than a collective effect for these nearly
rod-like molecules.  Collective effects should become more important if we
increase the interaction between the directions of the molecular
dipoles---either by including dipole-dipole interactions in our simulation
model or by increasing the bend angle $\theta$ to make the molecules more
oblique.

In addition to these results for the polarization, we also measure the
molecular tilt angle in the simulations.  The average tilt angle for the system
is extracted from the nematic order tensor $Q$ defined by Eq.~(\ref{nematic}).
From the eigenvector corresponding to the maximum eigenvalue of $Q$, we can
calculate the tilt angle away from the layer normal.  Using this technique, we
measure the tilt angle as a function of applied electric field in the
simulation for two temperatures, $k_B T=0.7$ and $k_B T=1.3$.  The results are
shown in Fig.~5b.  We observe that the tilt angle responds more sharply to the
applied field at lower temperature than at higher temperature; that is, the
electroclinic coefficient drops with increasing temperature.  This temperature
dependence is similar to the temperature dependence of the electric
susceptibility shown in Fig.~5a.  At high applied field, the tilt angle
saturates at about $19^\circ$ for both temperatures.  We note that the tilt
angle, in contrast with the polarization, is a collective effect rather than a
single-molecule effect in this simulation, as shown by the much longer
equilibration time for the tilt angle.  Thus, the saturated tilt angle of
$19^\circ$ is not simply related to the molecular geometry, but depends on the
collective order of many molecules whose transverse dipoles have been aligned
by the applied electric field.

When the molecules tilt under an electric field, the thickness of the smectic
layers shrinks.  Because the simulation cell is flexible, the $z$-dimension of
the cell also shrinks.  At the temperature $k_B T=0.7$, the $z$-dimension of
the cell changes from 36.7 at $E=0$ to 35.0 at $E=10$, which is a contraction
by a factor of 0.954.  This contraction is analogous to the change in the
smectic layer spacing under an electric field observed in x-ray diffraction
experiments~\cite{crawford}, and the contraction factor can be interpreted as
the cosine of an x-ray tilt angle of $17.5^\circ$.  This x-ray tilt angle is
somewhat smaller than the tilt angle of $19^\circ$ associated with the
eigenvectors of $Q$, which corresponds to the orientational ordering of the
molecular cores observed in optical experiments.

\subsection{Field-Induced Phase Transition}

We carried out further studies of the five-degree molecular system in a larger
temperature range, and located the SmA-nematic transition at approximately
$k_B T=3.0$.  Above that temperature, the SmA phase melts and the system is
stable as a nematic state, with low positional correlations (smectic order
parameter below 0.3) but with very high orientational order.  Figure~6a shows a
slice of the nematic system at $k_B T=3.3$, viewed from the $x$ direction.
When the temperature is lowered from $k_B T=3.3$ to $k_B T=2.1$, the system
returns to the SmA phase with clearly defined layers (smectic order parameter
of about 0.8).  This is evidence that the system has a stable and reversible
SmA-nematic phase transition.

Under a strong electric field, the nematic phase has a surprising behavior.  We
apply $E=10$ to the nematic system at $k_B T=3.3$, not far above the
nematic-SmA transition temperature.  The system regains a large smectic order
parameter and again forms clearly defined layers, as shown in Fig.~6b.  This
figure shows a slice of the system, with five layers in cross section.  Thus we
observe in this simulation a field-induced nematic-SmA phase transition.  In
experiments, electric-field-induced isotropic-nematic-smectic phase transitions
have been observed in thermotropic liquid crystals~\cite{durand}, and the
critical behavior of the field-induced molecular tilt near the nematic-SmA
transition has been investigated~\cite{patel}.  A good understanding of these
effects in simulation will contribute to a better understanding of
field-induced phase transitions in experiment.

\section{Discussion}

This simulation study shows that the molecular shape is very important for the
phase behavior of liquid crystals.  In the system with the $45^\circ$ molecular
bend angle, the steric repulsion based on molecular shape provides the driving
force for molecular tilt order in a SmC phase, even without intermolecular
dipole-dipole interactions.  In the system with the $5^\circ$ bend angle, the
molecules are closer to rigid rods, so they do not exhibit a SmC phase with
spontaneous tilt order.  Still, even a $5^\circ$ molecular bend leads to a
substantial electroclinic effect, which would be totally absent for rigid rods.
(Rigid rods with transverse electric dipoles would align their dipoles with an
applied electric field, but this alignment would not lead to any molecular
tilt.)  Preliminary simulation results for molecules with a $9^\circ$ bend
angle (not presented here) suggest that the phase transitions shift
dramatically from the $5^\circ$ molecules, confirming the influence of small
changes in molecular shape.  Hence, one conclusion of this study is that
collective intermolecular properties like molecular tilt and transition
temperatures are quite sensitive to slight details of molecular shape.  This
conclusion is somewhat disappointing from the perspective of modeling unique
properties of particular liquid-crystal compounds, as opposed to generic
properties based on molecular symmetry, because it implies that one must
describe the molecular structure very precisely in order to predict properties
like tilt and transition temperatures.

Another conclusion of this study is that the distribution of molecular tilts in
the SmA phase is more complex than is often supposed.  In the absence of an
applied electric field, the molecules do not stand up as rigid rods along the
layer normal.  Rather, there is disorder in the molecular tilt, with all of the
molecules tilting away from the layer normal in random azimuthal directions.
Some of this disorder takes the form of vortices in the tilt projected into the
smectic layer plane.  When an electric field is applied, it has two effects:
it increases the magnitude of the tilt angles and it increases the order in the
azimuthal direction of the tilt.  These two effects combine to give the
electroclinic tilt angle associated with the eigenvectors of the nematic order
tensor $Q$.  For that reason, this tilt angle is somewhat greater than the
x-ray tilt angle associated with the contraction of the smectic layers, which
arises only from the increase in the magnitude of the molecular tilt angles.
This result suggests that experimental measurements of the electroclinic effect
cannot be interpreted purely as tilting of rigid rods or as ordering of $xy$
spins, but rather as a combination of both.

The vortices observed in the SmA phase of the simulation are particularly
intriguing defects.  These vortices appear to be equivalent to the topological
defects that mediate the Kosterlitz-Thouless ordering transition in the
two-dimensional $xy$ model~\cite{xy}.  Thus, they suggest that the SmA phase is
analogous to the disordered phase of the $xy$ model and the SmC phase to the
ordered phase.  It is surprising that our three-dimensional simulation shows
point vortices that are uncoupled from one smectic layer to the next, and do
not thread through all five layers of the system.  This uncoupling presumably
occurs because, as noted earlier, there is very little interaction between the
tilt directions in adjacent layers due to the short-range repulsive potential
and the lack of interdigitation between layers.  The observation of these
defects leads to several questions for future research.  For example, how do
the defects evolve as a small electric field is applied?  In a system with a
SmA-SmC transition, what happens to the defects when the temperature drops
toward the transition?  Furthermore, if the interaction between molecules had a
longer range, would the point-like ``pancake'' vortices turn into vortex lines
as in conventional type II superconductors~\cite{crabtree}, or would they be
driven out of the system completely?  This final question is a key issue for
experimental systems in which the tilt directions of adjacent layers are
strongly coupled.

In summary, we have simulated smectic ordering in liquid crystals composed of
bent-rod molecules interacting through a soft repulsive potential.  The system
of highly bent molecules shows a SmC phase with spontaneous tilt ordering,
while the system of only slightly bent molecules shows a SmA phase with a
substantial induced tilt under an applied electric field.  These results show
the high sensitivity of molecular tilt ordering to the molecular shape, and
show the distribution of molecular tilt that controls the electroclinic effect.

\acknowledgments

This work was supported by the U. S. Navy Grant No.\ N00014-97-1-G003, the
National Science Foundation Grant No.\ DMR-9702234-1, and the Donors of the
Petroleum Research Fund, administered by the American Chemical Society.

\pagebreak

\epsfclipon

\begin{figure}
\centering\leavevmode\epsfxsize=3.2in\epsfbox{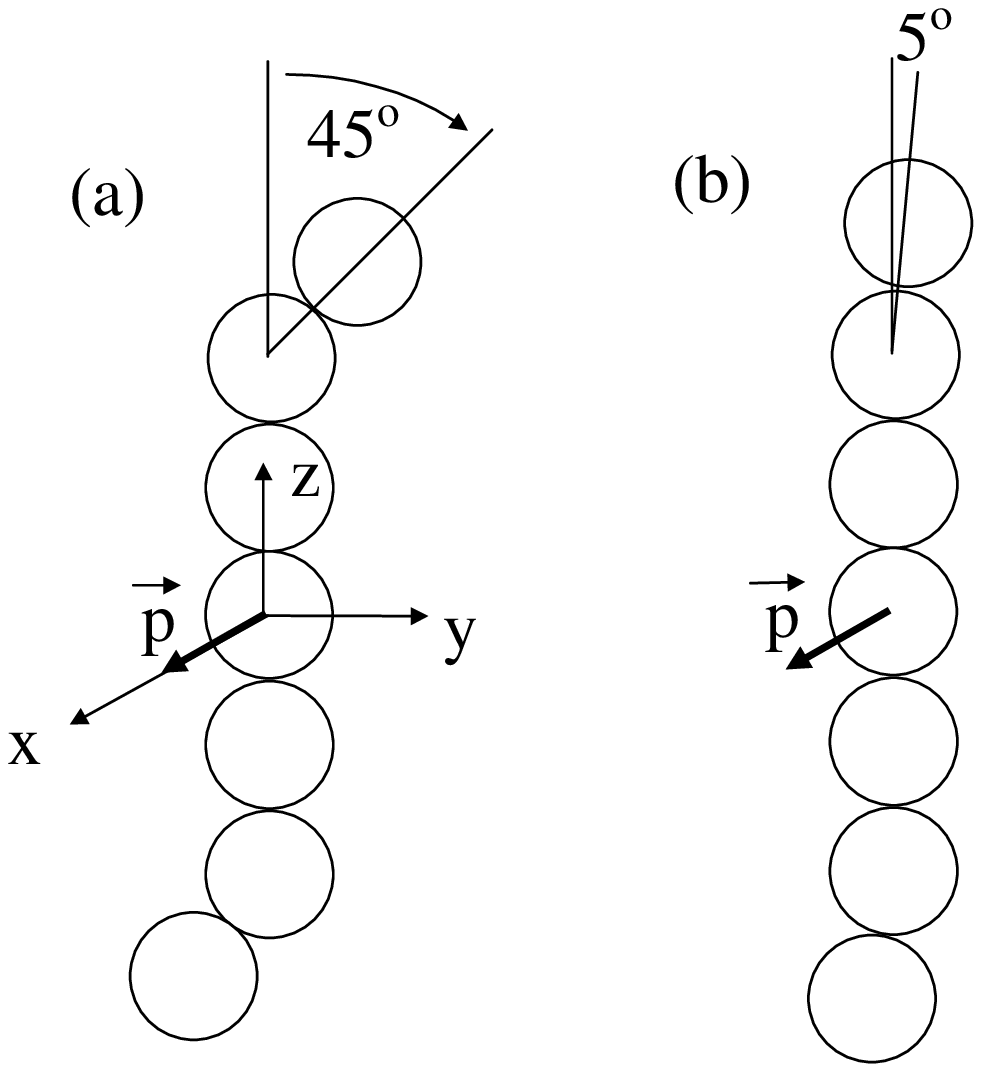}\bigskip
\caption{Basic molecular shape, with the bend angle $\theta$ between the core
and tail portions of the molecule.}
\end{figure}

\begin{figure}
\centering\leavevmode\epsfxsize=2.4in\epsfbox{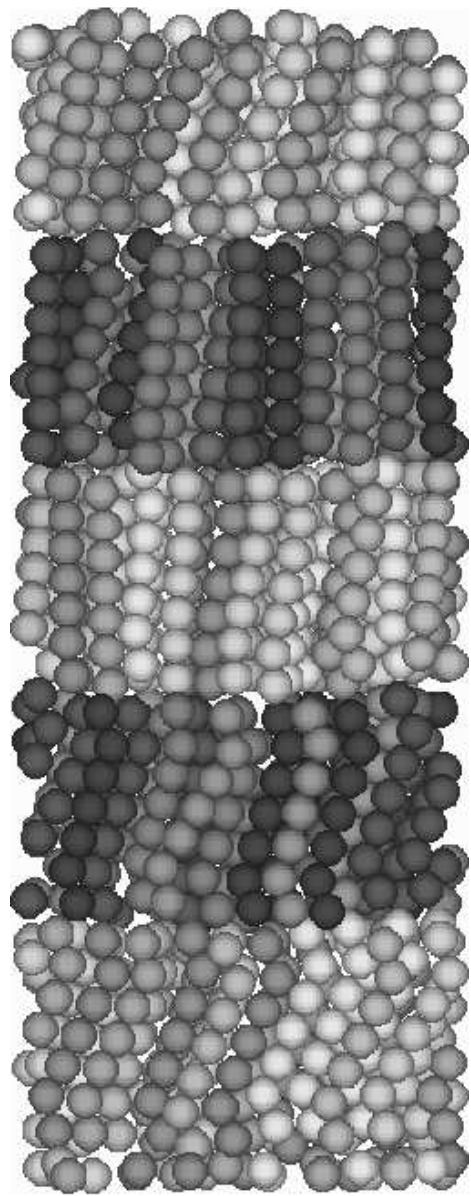}\bigskip
\caption{For the molecules with bend angle $\theta=45^\circ$, the simulations
show a SmC phase.  The direction of the molecular tilt varies from layer to
layer.  The molecules are drawn in different shades of gray in order to
distinguish them.}
\end{figure}

\begin{figure}
\centering\leavevmode(a)\epsfxsize=2.4in\epsfbox{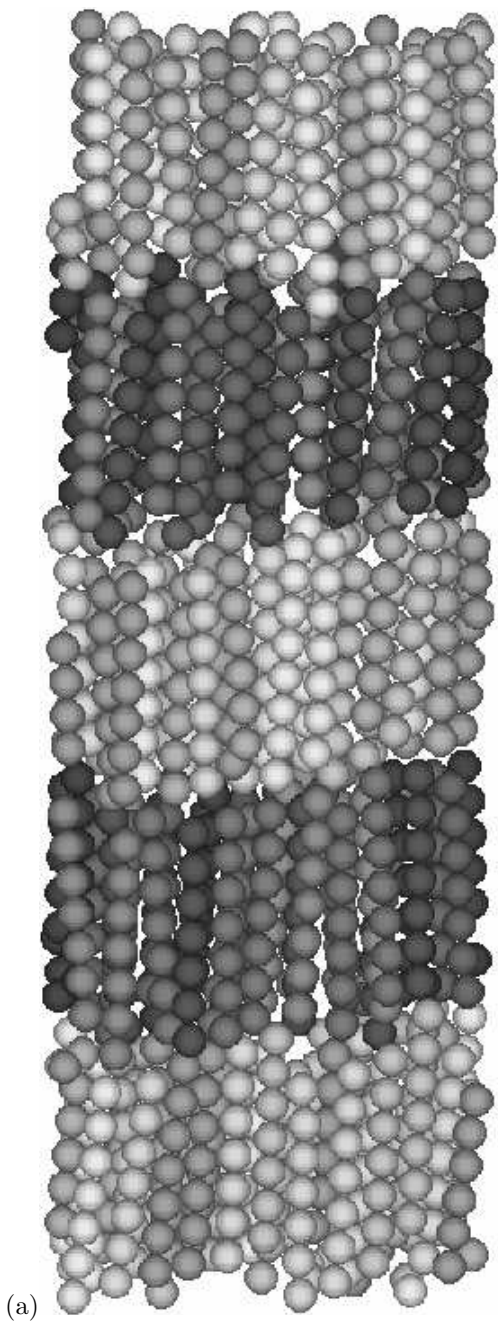}\bigskip

\centering\leavevmode(b)\epsfxsize=2.4in\epsfbox{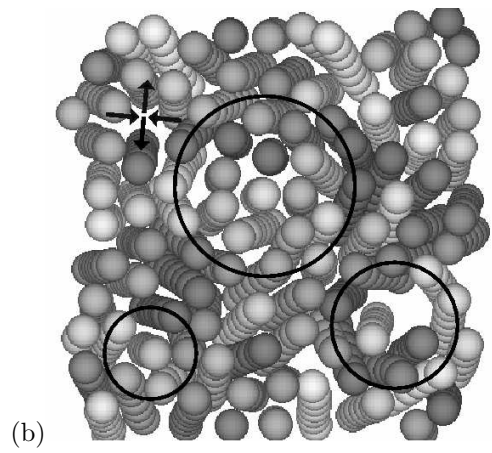}\bigskip
\caption{For the molecules with $\theta=5^\circ$, the simulations show a SmA
phase at $k_B T=1.3$.  (a)~Side view.  (b)~Top view of a single smectic layer.
The circles indicate three vortices with positive topological charge (two
right-handed and one left-handed), and the arrows indicate one vortex with
negative topological charge.}
\end{figure}

\begin{figure}
\centering\leavevmode(a)\epsfxsize=2.4in\epsfbox{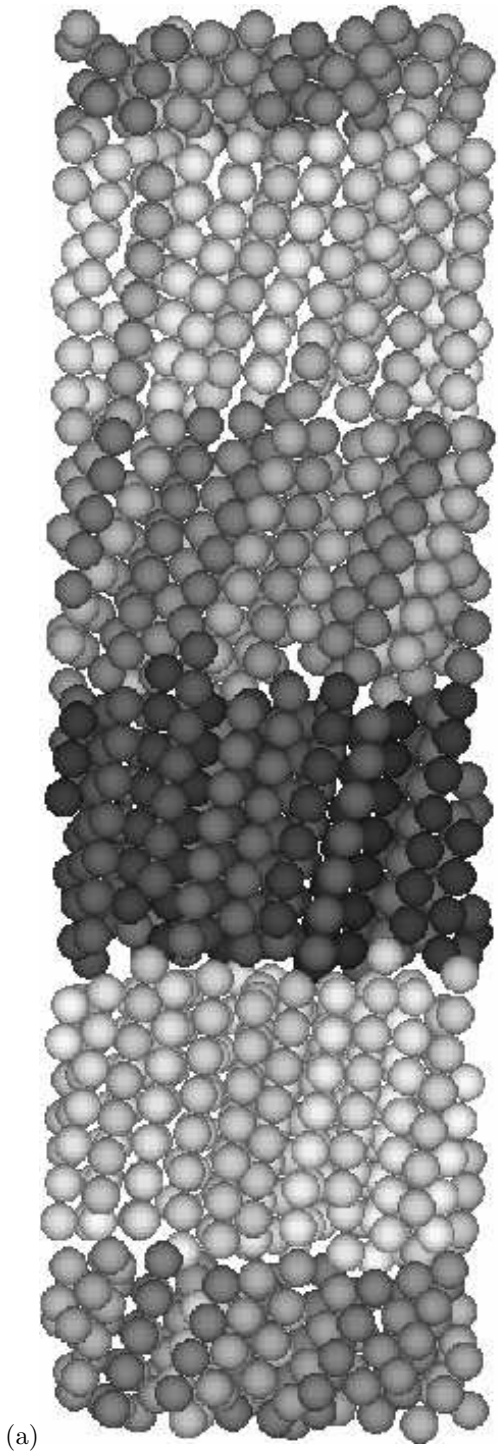}\bigskip

\centering\leavevmode(b)\epsfxsize=2.4in\epsfbox{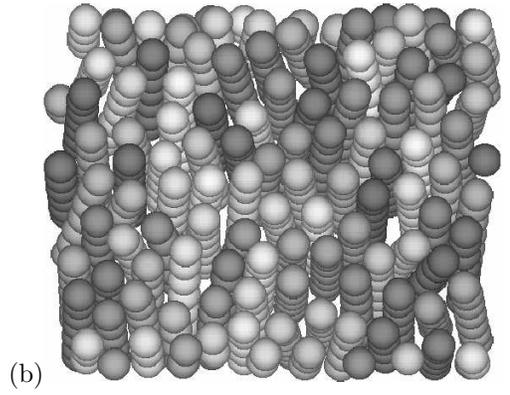}\bigskip
\caption{Under an applied electric field in the $x$ direction, the molecules
tilt with respect to the smectic layer normal, showing an electroclinic effect.
This picture shows the SmA system at $k_B T=1.3$ under a strong field $E=10$,
with an induced tilt of approximately $19^\circ$.  (a)~Side view.  (b)~Top view
of a single smectic layer, showing that the vortices have disappeared.}
\end{figure}

\begin{figure}
\centering\leavevmode(a)\epsfxsize=3.2in\epsfbox{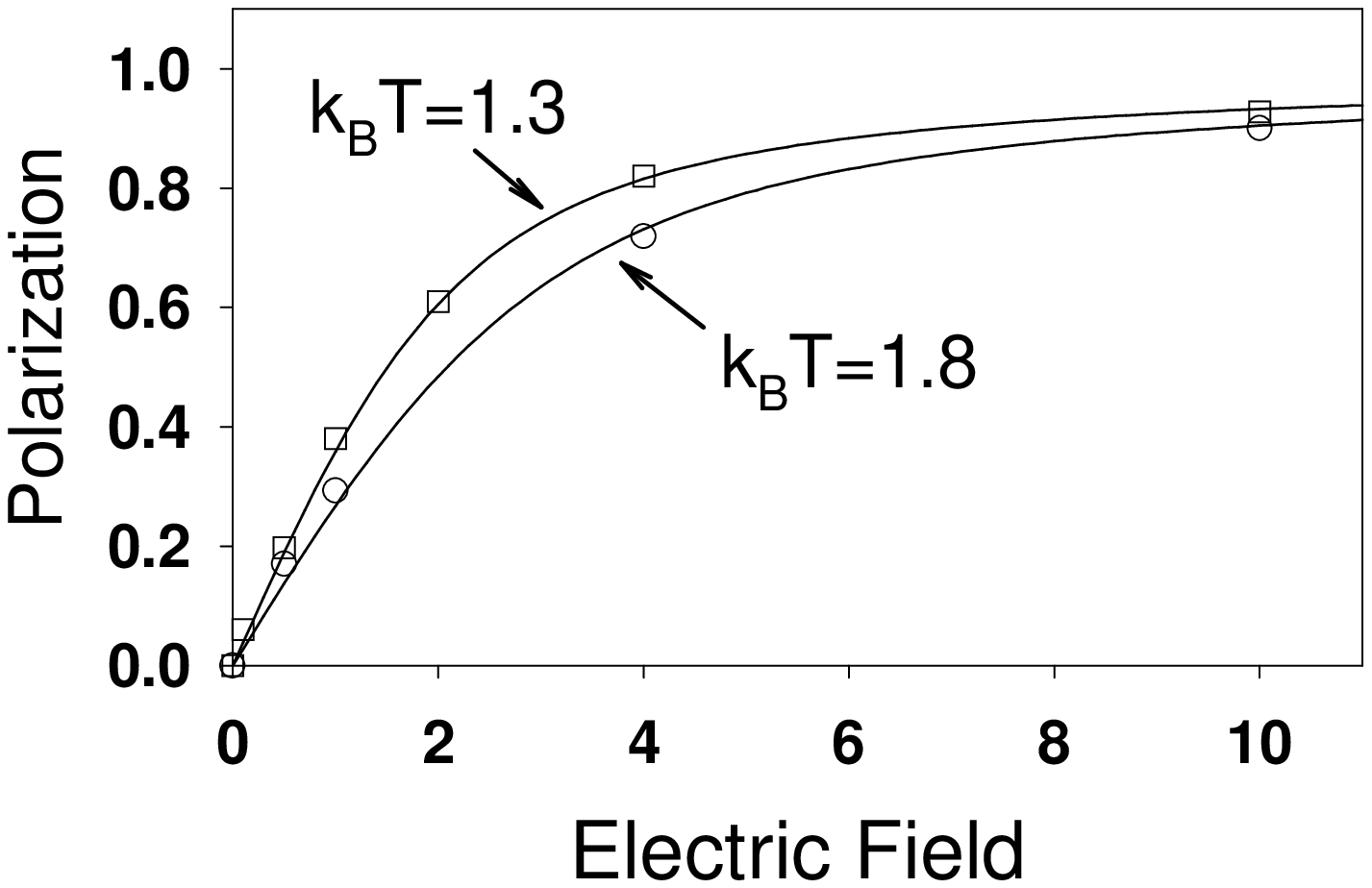}\bigskip

\centering\leavevmode(b)\epsfxsize=3.2in\epsfbox{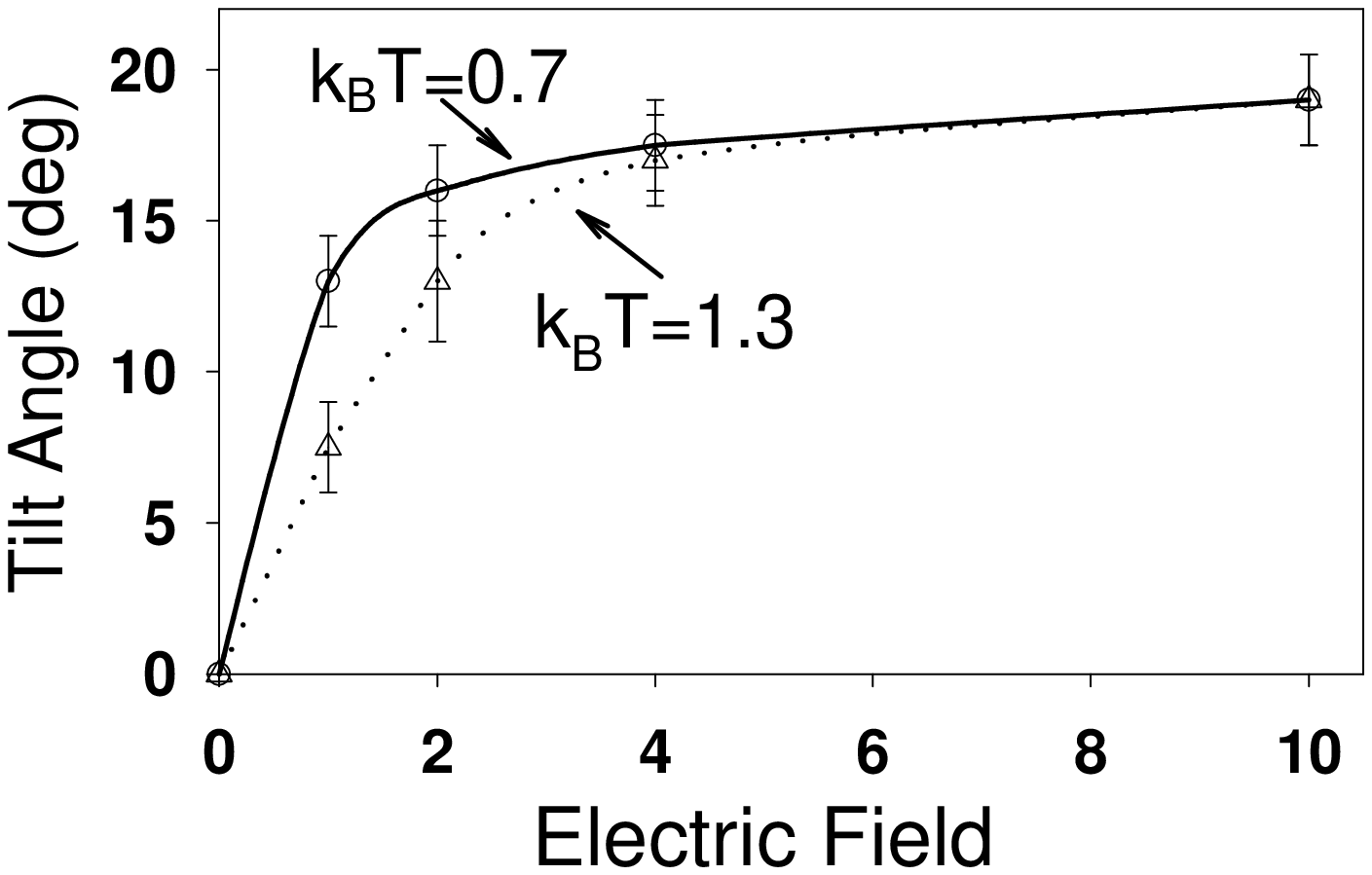}\bigskip
\caption{(a)~Induced polarization of the SmA phase as a function of applied
electric field, compared with the prediction of Eq.~(\protect\ref{polar}) from
a two-dimensional spin model.  Polarization is measured in units of the
molecular dipole moment $p$, temperature in units of the Lennard-Jones
parameter $\epsilon$, and electric field in units of $\epsilon/p$.
(b)~Induced tilt angle of the SmA phase as a function of applied electric
field.  In part b, the lines are guides to the eye.}
\end{figure}

\begin{figure}
\centering\leavevmode(a)\epsfxsize=2.4in\epsfbox{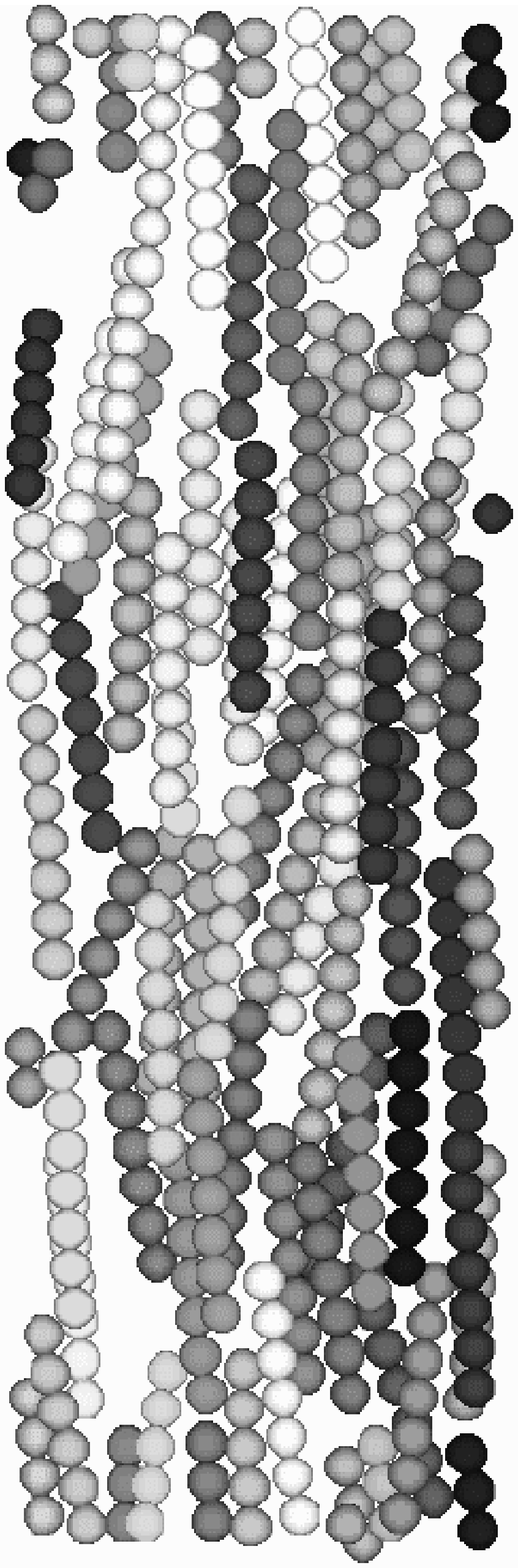}\bigskip

\centering\leavevmode(b)\epsfxsize=2.4in\epsfbox{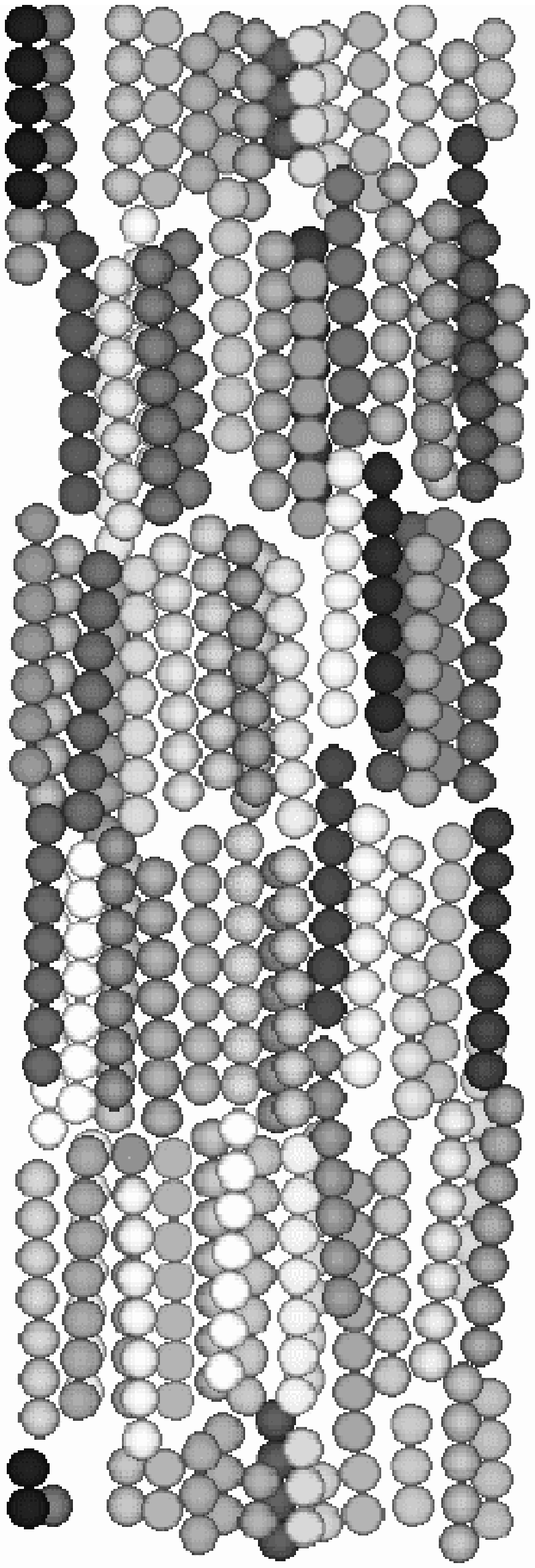}\bigskip
\caption{(a)~At $k_B T>3.0$, the SmA phase melts into a nematic phase with no
positional correlations but with very high orientational correlations.  This
image shows a side view of a slice through the nematic phase at $k_B T=3.3$.
(b)~Under a strong electric field, the nematic phase regains smectic order.
This image shows a side view of a slice through the system at $E=10$ and $k_B
T=3.3$.  Five smectic layers can be seen.}
\end{figure}


\begin{references}
\bibitem{meyer} R. B. Meyer, Mol. Cryst. Liq. Cryst. {\bf 40}, 33 (1977).
\bibitem{garoff} S. Garoff and R. B. Meyer, Phys. Rev. Lett. {\bf 38}, 848
(1977).
\bibitem{collings} N. Collings, W. A. Crossland, R. C. Chittick, and
M. F. Bone, Proc. SPIE Int. Soc. Opt. Eng. {\bf 963}, 46 (1989).
\bibitem{andersson} G. Andersson, I. Dahl, L. Komitov, S. T. Lagerwall,
K. Skarp, and B. Stebler, J. Appl. Phys. {\bf 66}, 4983 (1989).
\bibitem{blinc} T. Carlsson, B. \v{Z}ek\v{s}, C. Filipi\v{c}, A. Levstik, and
R. Blinc, Mol. Cryst. Liq. Cryst. {\bf 163}, 11 (1988); B. \v{Z}ek\v{s} and
R. Blinc, in {\it Ferroelectric Liquid Crystals} (Gordon and Breach,
Philadelphia, 1991), p. 365.
\bibitem{moddel} I. Abdulhalim and G. Moddel, Liquid Crystals {\bf 9}, 493
(1991).
\bibitem{devries} A. de Vries, A. Ekachai, and N. Spielberg, Mol. Cryst. Liq.
Cryst. Lett. {\bf 49}, 143 (1979); A. de Vries, Mol. Cryst. Liq. Cryst. Lett.
{\bf 49}, 179 (1979).
\bibitem{kn125} For reviews, see B. R. Ratna, G. P. Crawford, S. K. Prasad,
J. Naciri, P. Keller, and R. Shashidhar, Ferroelectrics {\bf 148}, 425 (1993);
G. P. Crawford, J. Naciri, R. Shashidhar, P. Keller, and B. R. Ratna, Mol.
Cryst. Liq. Cryst. A {\bf 263}, 223 (1995).  For the three-dimensional
structure of KN125, see A. Hermanns, C. Wilson, J. Patel, K. Gr\"{u}neberg,
K. Nelson, A. Townsend-Booth, and B. Ratna, Proc. SPIE Int. Soc. Opt. Eng.
{\bf 3297}, 73 (1998).
\bibitem{boulder} D. M. Walba, H. A. Razavi, A. Horiuchi, K. F. Eidman,
B. Otterholm, R. C. Haltiwanger, N. A. Clark, R. Shao, D. S. Parmar,
M. D. Wand, and R. T. Vohra, Ferroelectrics {\bf 113}, 21 (1991).
\bibitem{tarazona} A. M. Somoza and P. Tarazona, J. Chem. Phys. {\bf 91}, 517
(1989).
\bibitem{wca} J. D. Weeks, D. Chandler, and H. C. Andersen, J. Chem. Phys.
{\bf 54}, 5237 (1971); D. Chandler, J. D. Weeks, and H. C. Andersen, Science
{\bf 220}, 787 (1983).
\bibitem{affouard} F. Affouard, M. Kroger, and S. Hess, Phys. Rev. E {\bf 54},
5178 (1996).
\bibitem{fukuda} S. Inui, N. Iimura, T. Suzuki, H. Iwane, K. Miyachi,
Y. Takanishi, and A. Fukuda, J. Mater. Chem. {\bf 6}, 671 (1996).
\bibitem{rudquist} P. Rudquist, J. P. F. Lagerwall, M. Buivydas, F. Gouda,
S. T. Lagerwall, N. A. Clark, J. E. MacLennan, R. Shao, D. A. Coleman,
S. Bardon, T. Bellini, D. R. Link, G. Natale, M. A. Glaser, D. M. Walba,
M. D. Wand, and X.-H. Chen, J. Mater. Chem. {\bf 9}, 1257 (1999).
\bibitem{ohern} C. S. O'Hern, T. C. Lubensky, and J. Toner, preprint
cond-mat/9904415.
\bibitem{glaser} M. A. Glaser, R. Malzbender, N. A. Clark, and D. M. Walba, J.
Phys. Condens. Mat. {\bf 6}, A261 (1994).
\bibitem{xy} D. R. Nelson, in {\it Phase Transitions and Critical Phenomena},
Vol. 7, edited by C. Domb and J. L. Lebowitz (Academic Press, New York, 1983).
\bibitem{crabtree} G. W. Crabtree and D. R. Nelson, Phys. Today {\bf 50} (4),
38 (1997).
\bibitem{crawford} G. P. Crawford, R. E. Geer, J. Naciri, R. Shashidhar, and
B. R. Ratna, Appl. Phys. Lett. {\bf 65}, 2937 (1994).
\bibitem{durand} I. Lelidis and G. Durand, Phys. Rev. Lett. {\bf 73}, 672
(1994).
\bibitem{patel} S.-D. Lee and J. S. Patel, Phys. Lett. A {\bf 155}, 435 (1991).
\end{references}
\end{document}